\documentclass[twocolumn,prl,nobibnotes,superscriptaddress]{revtex4-2}

\usepackage{graphicx}
\usepackage{amsmath,amssymb,bm}
\usepackage{dcolumn}
\usepackage{bm}
\usepackage{color}
\usepackage{gensymb}
\definecolor{darkblue}{rgb}{0,0,.6}

\usepackage[
colorlinks=true,
citecolor=blue,
setpagesize=false]{hyperref}

\begin{document}

\title{X-ray magnetic circular dichroism and resonant inelastic X-ray scattering explained:\\
role of many-body correlation and valence fluctuations}
%role of many-body correlation and mixed-valence fluctuations}

%{Magnetic circular dichroism of X-ray absorption and resonant inelastic x-ray scattering in doped manganese oxides: Role of many-body correlation of intermediate core-valence states}

\author{Beom Hyun Kim} %\email{bomisu@ibs.re.kr}
\thanks{These authors contributed equally to this work}
\affiliation{Department of Physics and Astronomy, Seoul National University, Seoul 08826, Republic of Korea}
\affiliation{Center for Theoretical Physics of Complex Systems, Institute for Basic Science, Daejeon 34126, Republic of Korea}

\author{Sang-Jun Lee} %\email{sangjun2@slac.stanford.edu}
\thanks{These authors contributed equally to this work}
\affiliation{Stanford Synchrotron Radiation Lightsource, SLAC National Accelerator Laboratory, Menlo Park, California 94025, USA}

\author{H. Huang}
\affiliation{Stanford Synchrotron Radiation Lightsource, SLAC National Accelerator Laboratory, Menlo Park, California 94025, USA}
\affiliation{Department of Materials Science and Institute of Optoelectronics, Fudan University, Shanghai 200433, China}

\author{D. Lu}
\affiliation{Stanford Synchrotron Radiation Lightsource, SLAC National Accelerator Laboratory, Menlo Park, California 94025, USA}

\author{S. S. Hong}
\affiliation{Department of Materials Science and Engineering, University of California, Davis, CA 95616, USA}

\author{S. Lee}
\affiliation{Department of Physics and Materials Research Laboratory, University of Illinois, Urbana, Illinois 61801, USA}

\author{P. Abbamonte}
\affiliation{Department of Physics and Materials Research Laboratory, University of Illinois, Urbana, Illinois 61801, USA}

\author{Y. I. Joe}
\affiliation{National Institute of Standards and Technology, Boulder, CO 80305, USA}

\author{P. Szypryt}
\affiliation{National Institute of Standards and Technology, Boulder, CO 80305, USA}

\author{W. B. Doriese}
\affiliation{National Institute of Standards and Technology, Boulder, CO 80305, USA}

\author{D. S. Swetz}
\affiliation{National Institute of Standards and Technology, Boulder, CO 80305, USA}

\author{J. N. Ullom}
\affiliation{National Institute of Standards and Technology, Boulder, CO 80305, USA}

\author{C.-C. Kao}
\affiliation{SLAC National Accelerator Laboratory, Menlo Park, California 94025, USA}

\author{J.-S. Lee} \email{jslee@slac.stanford.edu}
\affiliation{Stanford Synchrotron Radiation Lightsource, SLAC National Accelerator Laboratory, Menlo Park, California 94025, USA}

\author{Bongjae Kim} \email{bongjae@knu.ac.kr}
\affiliation{Department of Physics, Kyungpook National University, Daegu 41566, Republic of Korea}

\date{\today}

\maketitle

\noindent
\textbf{
X-ray magnetic circular dichroism (XMCD) and resonant inelastic X-ray scattering with magnetic circular dichroism (RIXS-MCD) provide unparalleled insights into the electronic and magnetic dynamics of complex materials.
However, interpreting their spectra in mixed-valence systems remains challenging due to intricate many-body interactions and enhanced charge fluctuations. In this study, by utilizing the Anderson impurity model with a full consideration of charge transfer (CT), many-body core-valence exchange correlation (CVEC) effects, and Jahn-Teller (JT) distortions, we systematically investigate the XMCD and RIXS-MCD spectra for a prototypical mixed-valence ferromagnet, La$_{0.7}$Sr$_{0.3}$MnO$_3$ film. We demonstrate that simple calculation with limited CT effects fails to capture characteristic substructures observed experimentally. In contrast, an adequate treatment of CT and CVEC effects yields a more consistent description of both XMCD and RIXS-MCD spectra, providing practical guidance for the interpretation of dichroic x-ray spectroscopies in mixed-valence transition-metal oxides. Furthermore, we discuss the role of the JT effect in Mn$^{3+}$ ions in the determination of their spectra.}

\section{Introduction}

 Strongly correlated electron systems, where electronic, orbital, spin, and structural degrees of freedom interact in complex ways, have been a central focus of condensed matter research for decades~\cite{Coey1,Coey2,jin1994thousandfold,moritomo1996giant,kimura1996interplane,kimura2003magnetic,hur2004electric,bednorz1986possible,keimer2015quantum}. The intricate coupling of these degrees of freedom is responsible for the emergence of novel phases, such as the $J_{\rm eff}=1/2$ Mott state in iridates~\cite{kim2008novel,kim2009phase}, unconventional behavior in high-$T_c$ superconductors~\cite{oles2019orbital}, and the complex magnetic phase diagrams in perovskite heterostructures~\cite{Coey1,Coey2,lee2013titanium,salluzzo2009orbital}. Understanding the microscopic origins of these emergent properties requires disentangling the contributions of spin, orbital, and charge dynamics, which often interact in nontrivial ways, particularly in mixed-valence systems.

 Distinguishing between the various contributions (\emph{e.g.}, spin vs. orbital) is critical and experimentally challenging, especially in systems where these degrees of freedom are entangled. X-ray magnetic circular dichroism (XMCD) has long been a powerful tool for disentangling spin and orbital contributions in magnetic systems~\cite{van1986experimental,stohr1999exploring,chen1995experimental}. By exploiting the element-specific nature of XMCD, researchers can apply the sum rules to extract spin and orbital moment values from absorption spectra~\cite{chen1995experimental,thole1992x,carra1993x}.
 More recently, resonant inelastic X-ray scattering (RIXS), particularly in the form of RIXS-MCD, provides a complementary approach capable of probing both ground and excited state properties~\cite{Kotani2001,ament2011resonant,strange1991dichroic,duda1994magnetic,yablonskikh2001origin,braicovich1999magnetic,miyawaki2017compact,inami2017magnetic}. Furthermore, RIXS-MCD has been demonstrated to be applicable even in systems where time-reversal symmetry is preserved~\cite{Hariki2025}. It is expected that RIXS-MCD could reveal excitation spectra, including spin, orbital, and charge excitations, offering a more comprehensive picture of the complexities of complex correlations.

 Despite the capability of XMCD and RIXS, their analysis in mixed-valence systems is often complicated by strong many-body interactions. These complexities are mainly influenced by various many-body multiplet effects, including atomic multiplet, charge transfer (CT), and core-valence exchange correlation (CVEC)~\cite{deGroot2008}. For example, widely used XMCD sum rules often lead to significant errors in certain systems~\cite{obrien1994orbital, weller1995microscopic, wu1992enhanced, nakajima1999electron}. These errors arise not only from the poorly determined expectation value of the magnetic dipole operator, commonly referred to as $\left< T_z \right>$ term, in the sum rules calculation~\cite{wu1992enhanced,nakajima1999electron,wu1994limitation}, but also from the CVEC effect of final-state many-body correlations, which mixes spectral weight between $L_2$ and $L_3$ edges~\cite{Laan2004,Piamonteze2009,Piamonteze2009b}. Therefore, a comprehensive analysis of XMCD and RIXS requires many-body calculations that appropriately incorporate all these effects~\cite{Haverkort2016,Hariki2018,Wang2019,Hariki2020}.

%Here, we newly establish a comprehensive theoretical approach for understanding XMCD and RIXS-MCD spectral features. We employ the Anderson impurity model (AIM) and incorporate full consideration of charge transfer (CT) and many-body core-valence exchange correlation (CVEC) effects that account for local electronic correlations in both valence and core orbitals and charge fluctuations in the valence orbitals. We found that many-body CVEC plays a significant role in the misinterpretation of XMCD and RIXS-MCD spectra based on simple dipole transitions between occupied core and unoccupied valence states. Further, we reveal that this approach is especially beneficial for interpreting $3d$ transition metals, where XMCD sum-rule analyses can lead to significant errors due to the coupling of spin and orbital contributions across the $L_2$- and $L_3$-edges~\cite{obrien1994orbital, weller1995microscopic, wu1992enhanced, nakajima1999electron}.
 In this study, we revisit the role of many-body correlations and valence fluctuations in the XMCD and RIXS-MCD spectra of the mixed-valence system La$_{0.7}$Sr$_{0.3}$MnO$_3$ (LSMO), by establishing a comprehensive theoretical approach. We employ the Anderson impurity model (AIM) and incorporate full consideration of CT and many-body CVEC effects as well as the Jahn-Teller (JT) effect of Mn$^{3+}$ ions that account for local electronic correlations in both valence and core orbitals and charge fluctuations in the valence orbitals. XMCD analysis in mixed-valence systems often relies on a weighted sum of spectra from different ions, here, Mn$^{3+}$ and Mn$^{4+}$ ions, which are calculated with restricted CT channels (e.g., single-electron transfer effect). However, this simplified approach frequently fails to reproduce some key characteristic spectral features, particularly at the $L_2$-edge~\cite{Kuepper2012}. Here, we demonstrate that an adequate consideration of full CT and many-body CVEC in the weighted-sum calculation is essential to describe both XMCD and RIXS-MCD spectra of LSMO. We analyze the microscopic transition channels for each $L_2$ and $L_3$ edges, and identify the contributions from different dipole transitions. Consequently, our combined theoretical and experimental approach provides deeper insights into XMCD and RIXS-MCD processes, particularly for light $3d$ transition-metal systems with mixed-valance, offering a complete and robust framework for studying other correlated materials.

%===========================================
\begin{figure*}[!th]
\centering
\includegraphics[width=0.95\textwidth]{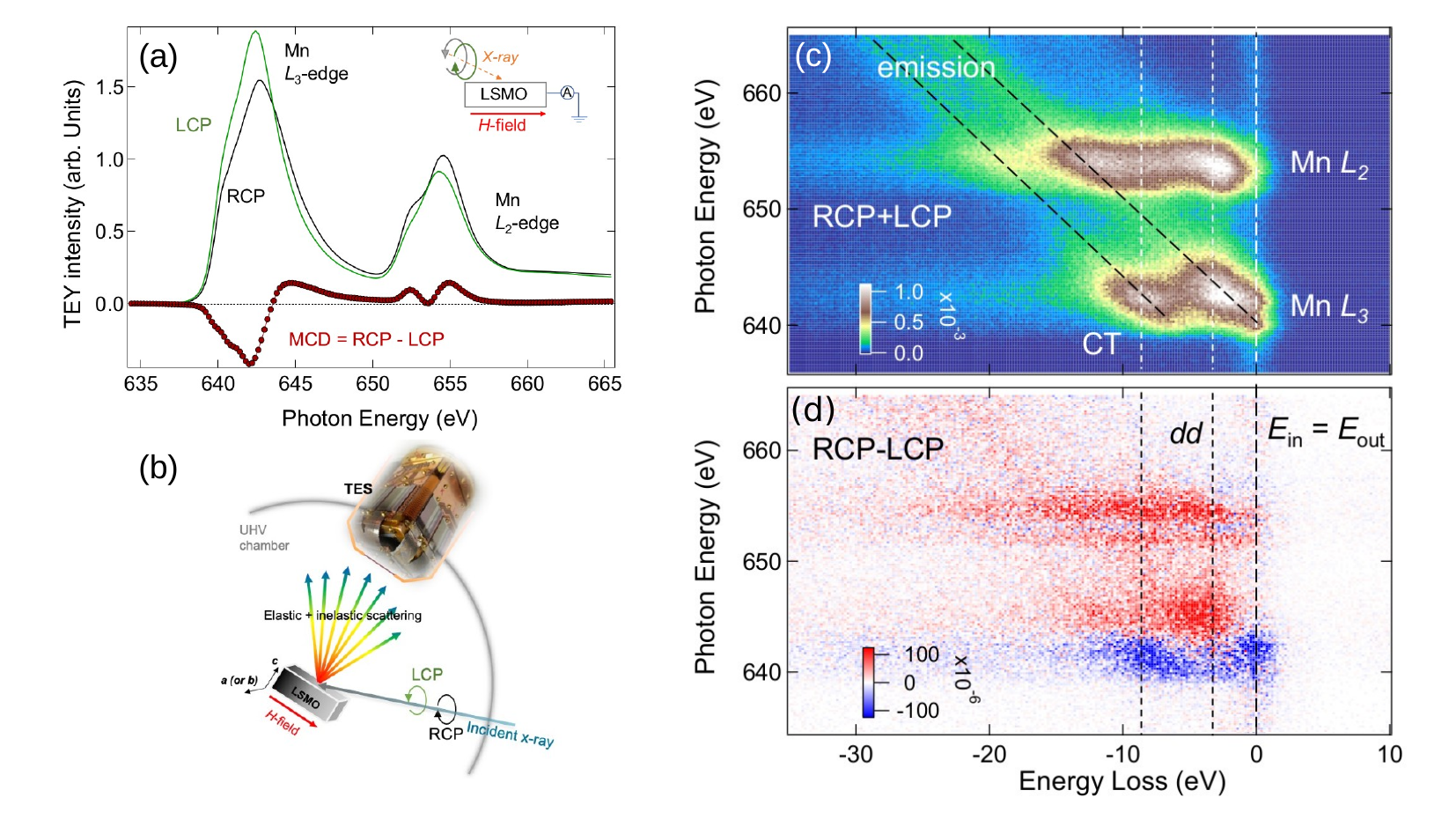}
\caption {
 (a) X-ray absorption spectroscopy (XAS) and magnetic circular dichroism (MCD) spectra of La$_{0.7}$Sr$_{0.3}$MnO$_3$ (LSMO)  measured in the total electron yield (TEY) mode.
 The black and green curves are the absorption spectra of right circular polarization (RCP) $\mu_+$ and left circular polarization (LCP) $\mu_-$,  respectively.
 The curve with circle symbols is the MCD spectrum ($\Delta \mu$).
 The inset in (a) shows the experimental configuration for the MCD measurement. The symbol \textcircled{A} denotes an ammeter to measure the TEY.
 (b) Schematic drawing of the resonant inelastic X-ray scattering (RIXS) measurement set-up using a transition-edge-sensor (TES) spectrometer.
 (c) A RIXS map as functions of the energy loss ($\Delta E = E_{\rm out} - E_{\rm in}$) and the photon energy ($E_{\rm in}$) of the LSMO measured by the TES spectrometer.
The map represents the sum of RIXS intensities for the RCP and LCP incident photons. (d) A RIXS-MCD map obtained by differing two RIXS maps measured with RCP and LCP incident photons.
Along the dashed line around  $-9 < \Delta E < -3$ eV, a strong dichroic intensity was observed at $E_{\rm in} \sim 644.6$ eV, which corresponds to the $dd$-excitation.
}
\label{Fig1}
\end{figure*}
%===========================================

\section{Results}

\subsection{Experimental results}

We perform the X-ray absorption spectroscopy (XAS) of epitaxial LSMO films with two polarizations, right circular polarization (RCP) and left circular polarization (LCP) (see the detail in Methods section). The XMCD spectrum ($\Delta\mu$) is generated as the difference of the two XAS spectra obtained with RCP ($\mu_+$) and LCP ($\mu_+$). As shown in Fig.~\ref{Fig1}(a),
obtained spectra are well consistent with the previous reports~\cite{yu2010interface,lee2010hidden,Koide2001,deJong2005,Aruta2009,Shibata2014,Shibata2018}.
At the $L_3$-edge, $\Delta \mu$ exhibits two dichroism features at around 642.1 eV and 644.6 eV, with a 2.5 eV energy difference.
Similarly, at the $L_2$-edge, $\Delta \mu$ displays two features at around 652.5 eV and 655.0 eV, indicating the same energy difference.
This correlation suggests a connection between the dichroism features of the $L_3$-edge and those of the $L_2$-edge, which becomes uncovered later.

Considering the magnetic exchange splitting of the $3d$ valence band between up and down spins, the sign of XMCD between the $L_3$ and $L_2$ edges is typically opposite~\cite{van1986experimental,stohr1995determination,chen1995experimental,thole1992x,carra1993x}, as observed in the two main XMCD peaks at 642.1 eV for the $L_3$ edge and 652.5 eV for the $L_2$ edge (see Fig.\ref{Fig1}(a)).
However, the other subpeak at 644.6 eV eV for the $L_3$ edge and 655.0 eV for the $L_2$ edge exhibit the same sign of XMCD.
This discrepancy, along with the overshoot observed at the $L_3$-edge XMCD resembling the diffused moment behavior reported in Ref.~\onlinecite{obrien1994orbital}, leads to a strong overestimation and underestimation of the orbital and spin moment, respectively, when applying the sum-rules. This representatively shows the limitations of XMCD sum-rules~\cite{wu1994limitation,Laan2004,Piamonteze2009,Piamonteze2009b}, particularly in exploring orbital angular momentum, and raises fundamental questions on the application of the sum-rules.
Understanding the origin of this sign-reversal behavior and identifying the legitimate regime of the sum-rules is essential.
A comprehensive analysis of XMCD spectra calls for microscopic theories to determine the complex multiplet structure of Mn $3d$ orbitals.

We further explore RIXS and RIXS-MCD spectra (see the detail in Methods section). By monitoring $\Delta E$ for each $E_{\rm in}$ through RIXS measurement (see Fig.~\ref{Fig1}(b) and (c)), we generate a map of LSMO as functions of $\Delta E$ and $E_{\rm in}$ by summing the maps measured with RCP and LCP at $T = 25$ K.
Figure~\ref{Fig1}(c) is a magnified view of the RIXS map over the Mn $L$-edge region, where we identify quasi-elastic, $dd$-excitation, CT, and fluorescence signals. The quasi-elastic ($\Delta E \sim  0$) includes not only the elastic but also phonon and magnon contributions (typically on the energy scale of tens to hundreds of meV) due to the spectrometer's energy resolution of 1.5 eV~\cite{ament2011resonant,ghiringhelli2004low,ghiringhelli2006resonant}. Mn's $dd$-excitation ($\Delta E \sim -3$ eV) and CT ($-10$ eV $< \Delta E < -3$ eV) are well pronounced (see the theoretical analysis in next section). The fluorescence appears along the diagonal direction guided with dashed lines in Fig.~\ref{Fig1}(c).

The RIXS-MCD map, obtained from the difference of two RIXS maps with RCP and LCP incident photons, is presented in Fig.~\ref{Fig1}(d). We observe the strong dependency of the RIXS-MCD map on the incident photon energy, showing the separation of RCP and LCP dominant spectra. Positive and negative RIXS-MCD signals are extended over the loss energy when the incident energy is at around the XMCD absorption peaks. It supports that the overall fluorescence features of XMCD correspond well to the XMCD features measured using the total electron yield (TEY) mode at the Mn $L_{2,3}$ edges.

Given that the Fermi-level spin polarization of LSMO is highly pronounced~\cite{jonker1950ferromagnetic,tokura1994giant,park1998direct}, both elastic and inelastic processes are influenced by its magnetically polarized behavior.
Interestingly, we observe that the XMCD feature at $E_{\rm in} \sim 644.6$ eV is more prominent in the RIXS-MCD compared to other excitations, which still correspond to the total MCD signal based on TEY measurements.
The energy loss value of this feature suggests that its origin is Mn’s $dd$-excitation, opening up possibilities for further discussion on the orbital moment investigated using the RIXS-MCD approach.

%===========================================
\begin{figure}[!t]
\centering
\includegraphics[width=0.95\columnwidth]{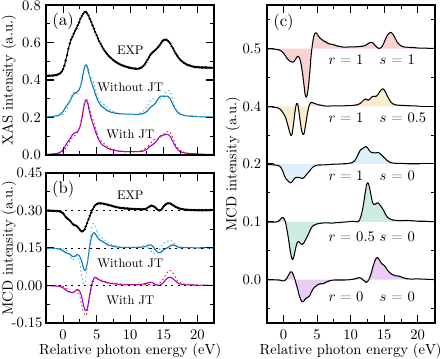}
\caption {
Theoretical spectra of (a) X-ray absorption spectroscopy (XAS) and (b) magnetic circular dichroism (MCD) of LSMO calculated by adding and subtracting the absorption spectra with right and left circular polarizations.
The spectra are obtained by the weighted sum of separate calculations of Mn$^{4+}$ and Mn$^{3+}$.
The solid magenta and blue lines correspond to the results from our model with and without the Jahn-Teller distortion, respectively. These calculations include full charge transfer (CT) states with configurations ranging from $d^{n+1}L^1$ to $d^{10}{L}^{10-n}$ ($n=3$ for Mn$^{4+}$ and $n=4$ for Mn$^{3+}$), where $L$ represents a ligand (bath) hole orbital.
For the comparison, the XAS and XMCD spectra calculated using the restricted CT states (limited to the $d^{n+1}L^1$ configuration), are shown as dotted lines.
 (c) The XMCD results for various CVEC strengths and hoping integrals between valence $3d$ and ligand (bath) orbitals.
 The Slater-Condon parameters are scaled by $s$ as $F_{pd}^2= 5.7508\times s$, $G_{pd}^1= 4.2864\times s$, and $G_{pd}^3 = 2.4382\times s$ eV, and the scale parameter $r$ is defined as $V_{pd\sigma}=-2 \times r$ and $V_{pd\pi}= 1 \times r$ eV. Other parameters are presented in Table~\ref{tb_para}.
}
\label{Fig_mcd_th}
\end{figure}
%===========================================

\subsection{Theoretical XMCD spectra}

As shown in Figs.~\ref{Fig_mcd_th}(a) and (b), we obtain the theoretical XAS and XMCD spectra which are highly consistent with experimental ones (see the detailed theoretical calculation in Methods section). To ensure comprehensive understanding, we investigate the XMCD behaviors focusing on the three key effects: CT, JT distortion, and CVEC.

Charge fluctuations in mixed-valence systems can be attributed not only to local CT effects via the covalency between Mn $3d$ and O $2p$ orbitals, but also to valence fluctuations over several Mn sites arising from a non-integer nominal valency. While the latter corresponds to inter-site cross-talk, the former induces localized charge fluctuations.
In the CI calculation, due to the computational cost, the CT states are frequently restricted to only the major $d^{n+1}L^1$ configuration~\cite{deGroot2005,stavitski2010ctm4xas}.
The spectrum of a mixed-valence system is then obtained from the spectra of each ionic state separately -- Mn$^{3+}$ and Mn$^{4+}$ for LSMO. Then, one simply performs the weighted sum of two spectra upon their mixing ratio.
This treatment successfully depict the main features of the experimental XMCD spectra in many cases~\cite{Lee2008,Lee2009,Kuepper2012,Cuartero2016,Backes2024}. However, these procedures fail to simulate the experimental XMCD spectra of highly metallic LSMO. Specifically, two important features, the extended subpeak spectra in the $L_3$-edge for the relative photon energy of around $0$ -- $2$ eV, and positive peak in the $L_2$-edge for the $13$ -- $14$ eV regime, are not accounted in the simple CT consideration with $d^{n+1}L^1$ configuration only (See the dotted data of Fig.~\ref{Fig_mcd_th}(b)).

In our AIM model, we treat the charge fluctuations of CT states more rigorously, since the CT effects are explicitly captured through strong hopping combined with a small CT energy.
Full consideration of CT of all orders enables us to capture simultaneously the valence fluctuations of Mn $d$ electrons ranging from $d^n$ to $d^{10}$ ($n=3,4$). Hence, our ground multiplet state well accounts for the CT nature beyond $d^{n+1}L^1$. Calculated ground multiplet state is composed of approximately 27\% $d^3 L^0$, 49\% $d^4 L^1$, 21\% $d^5 L^2$, and 3\% $d^6L^3$ configurations for Mn$^{4+}$, and 32\% $d^4 L^0$, 50\% $d^5 L^1$, 16\% $d^6 L^2$, and 2\% $d^7 L^3$ configurations for Mn$^{3+}$, demonstrating significant contribution of CT beyond $d^{n+1}L^1$ (see Supplementary Note 1). For LSMO, to properly describe the XMCD spectra, CT states up to at least $d^{n+3}L^3$ must be included (see Supplementary Note 5). Because dynamic charge fluctuations are more rigorously treated in our model than in partial CT approaches, we can successfully reproduce the key features of the experimental XAS and XMCD spectra. As shown in Fig.~\ref{Fig_mcd_th}(a) and (b), we clearly identify the more extended subpeak structure at $L_3$ and good account of the positive double-peak structure at $L_2$ XMCD spectra. These results underscore the importance of the full CT treatment, hence, the proper inclusion of the mixed-valence nature of Mn ions, for the comprehensively understanding  the core-level spectra of metallic LSMO.

 We further investigate the role of the Jahn-Teller (JT) distortion. While its role is less prominent for XAS, JT distortion plays a crucial role in shaping the fine structures of the XMCD as shown in Fig.~\ref{Fig_mcd_th}(b), particularly at the $L_2$ edge. Without the JT effect, between two peaks, a negative signal emerges at a relative photon energy of about 14 eV. Only when the $C$-type JT order explicitly is considered ($\theta_{JT} = 2\pi/3$ or $4\pi/3$), the characteristic positive two-peak structure is robustly reproduced in the calculated XMCD spectra. This also provides the strong evidence that the JT effect persists even in the metallic phases of LSMO. See Supplementary Note 6 for the detailed analysis of the JT effects.

%===========================================
\begin{figure*}[!t]
\centering
\includegraphics[width=0.95\textwidth]{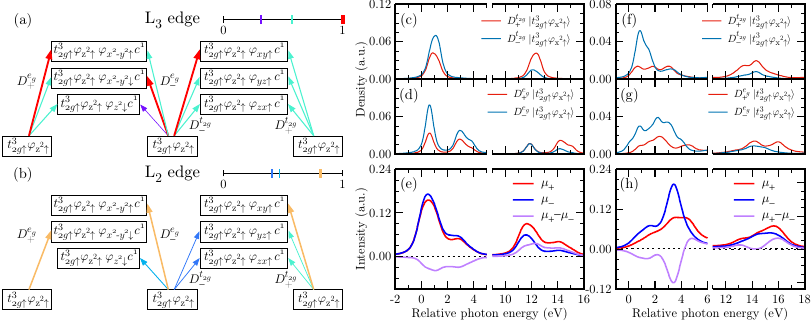}
\caption {
Schematic diagrams of dipole transitions from the $\left|t_{2g\uparrow}^3 \varphi_{z^2\uparrow} \right>$ state at the (a) $L_3$ and (b) $L_2$ edges. $D^{t_{2g}}_{+}$ and $D^{e_g}_{+}$ ($D^{t_{2g}}_{-}$ and $D^{e_g}_{-}$) denote the dipole transition operators for right (left) circularly polarized photons from core $2p$ to valence $t_{2g}$ and $e_{g}$ orbitals, respectively.
The thickness of arrows indicates the strength of dipole transitions, with relative values marked at the top-right horizontal lines.
The partial excitation distribution (PED) behaviors of multiplet states derived from $\big|t^3_{2g\uparrow} \varphi_{x^2\uparrow} \big>$,
which gives dominant contribution in ground state for $\theta_{JT}=2\pi/3$,
by the dipole operators $D_{+}^{t_{2g}}$, $D_{-}^{t_{2g}}$, $D_{+}^{e_{g}}$, and $D_{-}^{e_{g}}$ when the scale parameter $s$ is $0$ for (c) \& (d), and $1$ for (f) \& (g).
 The theoretical $L_3$-edge X-ray magnetic circular dichroism (XMCD) spectra for (e) $s = 0$ and (h) $s=1$.
 The scale parameter is defined as $F_{pd}^2= 5.7508\times s$, $G_{pd}^1= 4.2864\times s$, and $G_{pd}^3 = 2.4382\times s$ eV.
 Other parameters are presented in Table~\ref{tb_para}.
}
\label{Fig_dst}
\end{figure*}
%===========================================

 Now we include many-body correlation of core-valence states and systematically investigate the XMCD spectral evolution as functions of the CT and CVEC effects together. To this end, here, we adjust the hopping strengths and Slater-Condon parameters of core-valence orbitals using two scale parameters $r$ and $s$, such that $V_{pd\sigma}=-2 \times r$ and $V_{pd\pi}= 1 \times r$ eV, and $F_{pd}^2= 5.7508\times s$, $G_{pd}^1= 4.2864\times s$ and $G_{pd}^3 = 2.4382\times s$ eV. Hence, $r$ and $s$ act as a scaling factor for each CT and CVEC channel (Fig.~\ref{Fig_mcd_th}(c)). When both effects are absent ($r=0, s=0$), the $L_3$-edge XMCD spectra consist of a main peak with a negative sign and a subpeak with a positive sign, and corresponding $L_2$-edge has the opposite composition, a main peak with a positive and subpeak with a negative sign as expected ~\cite{van1986experimental,stohr1995determination,chen1995experimental,thole1992x,carra1993x}. With excluding the CVEC effect ($s=0$), we vary the strength of the CT channel. As we increase the hopping strengths ($r=0.5$), the positive (negative) subpeak in the $L_3$-edge ($L_2$-edge) progressively diminishes, and the negative (positive) main peak bifurcates to give two-peak structures. When full CT effects are included ($r=1$), the subpeak signatures are all gone, and we observe two negative peaks in the $L_3$-edge and two positive peaks in the $L_2$-edge, both evolved from the main peak in the ($r=0, s=0$) case (Fig.~\ref{Fig_mcd_th}(c)).

 The fine structure of XMCD peaks for the $r = 0, s = 0$ case depends solely on the \emph{local} correlation effects of Mn-$3d$ orbitals, such as crystal field energy and exchange Coulomb interactions, characterized by $F^2_{dd}$ and $F^4_{dd}$. Upon increasing $r$, the fine structure smoothed out. This indicates that strong $p$-$d$ hopping, which is due to the metallic nature of Mn ions, effectively screens the detailed local effects. In the absence of the CVEC, the $L_3$-edge ($L_2$-edge) XAS spectra can be directly interpreted with the simple dipole transition picture: excitations from the occupied core $p_{3/2}$ ($p_{1/2}$) orbitals to the unoccupied valence $3d$ orbitals.

%Now we discuss the microscopy of the transition channels.
Given that the spin-polarized ground state predominantly comprises the multiplet state with the $t_{2g\uparrow}^3e_{g\uparrow}^1$ and $t_{2g\uparrow}^3$ configurations for Mn$^{3+}$ and Mn$^{4+}$, respectively, we investigate its dipole transition behaviors. From the dipole selection rules, the transition from $p_{3/2,+3/2}$ ($p_{3/2,-3/2}$) to $\varphi_{x^2-y^2\uparrow}$ ($\varphi_{x^2-y^2\downarrow}$) orbitals of RCP (LCP) photons exhibits maximum amplitude at the $L_3$ edge. Conversely, the transition from $p_{1/2,+1/2}$ ($p_{1/2,-1/2}$) to $\varphi_{x^2-y^2\downarrow}$ ($\varphi_{x^2-y^2\uparrow}$) orbitals shows maximum amplitude at the $L_2$ edge, as illustrated in Fig.~\ref{Fig_dst}(a) and (b). (See Supplementary Fig. S6(a) and (b) for Mn$^{4+}$) case. Additionally, the transition from $p_{3/2,-3/2}$ to $\varphi_{xy\downarrow}$ orbitals of LCP photons contributes more significantly than the transition from $p_{3/2,+1/2}$ to $\varphi_{xy\downarrow}$ orbitals of RCP photons at the $L_3$ edge. Moreover, the transition from $p_{1/2,\pm 1/2}$ to $\varphi_{xy\downarrow}$ orbitals of LCP photons is forbidden at the $L_2$ edge. These spin and orbital dependencies of dipole transitions lead to the observed dichroism in Mn ions.

 To identify the microscopic contributions of each channels, we calculate the partial excitation density (PED) of the final states derived from $\big| t_{2g\uparrow}^3 \varphi_{x^2\uparrow} \big>$ through dipole transitions from core $2p$ to valence $t_{2g}$ and $e_{g}$ orbitals, $D_{\pm}^{t_{2g}}\big| t_{2g\uparrow}^3 \varphi_{x^2\uparrow} \big>$ and $D_{\pm}^{e_{g}}\big| t_{2g\uparrow}^3 \varphi_{x^2\uparrow}\big>$, which are the most dominant contributions in the ground multiplet state of Mn$^{3+}$ ions for $\theta_{JT} = 2\pi/3$. Here, we included fully included the hopping $r = 0$. The detailed calculations can be found in Supplementary Note 8.

 In Fig.~\ref{Fig_dst}(c) and (d), we display the calculated PED of the final states without CVEC effect ($r = 0$). The peaks of PED for $D_{\pm}^{t_{2g}}\big| t_{2g\uparrow}^3 \varphi_{x^2\uparrow}\big>$ appear at the eigenenergies of the final multiplet states, which consist of $t^3_{2g\uparrow} \varphi_{x^2\uparrow} t^1_{2g\downarrow}c^1$ configurations for $D_{\pm}^{t_{2g}}\big| t_{2g\uparrow}^3 \varphi_{x^2\uparrow}\big>$, and $t^3_{2g\uparrow}e^2_{g\uparrow}c^1$ and  $t^3_{2g\uparrow} \varphi_{x^2\uparrow} e^1_{g\downarrow} c^1$ configurations for $D_{\pm}^{e_{g}}\big| t_{2g\uparrow}^3 \varphi_{x^2\uparrow} \big>$. Here, $c$ denotes the core $2p$ hole. For $r=1, s=0$, the multiplet states with $t^3_{2g\uparrow} \varphi_{x^2\uparrow} t^1_{2g\downarrow} c^1$ configuration exhibit peaks at around 1 eV at the $L_3$ edge and 12.2 eV at the $L_2$ edge of relative photon energy. Additionally, the multiplet states with $t^3_{2g\uparrow}e^2_{g\uparrow}c^1$ and $t^3_{2g\uparrow} \varphi_{x^2\uparrow} e^1_{g\downarrow} c^1$ configurations show main peaks at around 0.6 eV and 3 eV at the $L_3$ edge, and 11.8 eV and 14.2 eV at the $L_2$ edge, respectively. Thus, the negative (positive) XMCD peak at around 1 eV (12.2 eV) at the $L_3$ ($L_2$) edge can be attributed to the contribution of multiplet states with $t^3_{2g\uparrow}\varphi_{x^2\uparrow} t^1_{2g\downarrow} c^1$ derived from the dipole transition from $p_{3/2}$ ($p_{1/2}$) to $t_{2g\downarrow}$. The negative (positive) XMCD peak at around 3 eV (14 eV) results from the sum of positive (negative) and negative (positive) contributions of the dipole transition from $p_{3/2}$ ($p_{1/2}$) to $e_{g \uparrow}$ or $e_{g \downarrow}$. Similar analysis is also accessible for $\big| t_{2g\uparrow}^3 \big>$ for Mn$^{4+}$ calculation as shown in Supplementary Fig. S6(c) and (d).

 The two-peak structures for $r=1, s = 0$ is a reflection of the electronic structures, where the large and small peak corresponds to the unoccupied Mn-$3d$ bands at around 1 and 3 eV, respectively (see Supplementary Note 2). The former is the mixture of the $t_{2g}$ and $e_g$, but the latter is mainly of the $e_g$ character. Note that the conventional sum-rule analysis, which separately treats $L_2$- and $L_3$-edge with \emph{circa} one-body picture, remains valid regardless of the CT effect as long as the many-body CVEC are absent.

 Now we discuss XMCD spectra. Before the microscopic PED analysis with CVEC ($s=1$), we explore the behaviors of the XMCD first upon the inclusion of the CVEC, displayed in Fig.~\ref{Fig_mcd_th} (e) and (f). Most importantly, the response of the $L_3$- and $L_2$-edge XMCD spectra are totally different upon the CVEC effect. As $s$ is increased from $0$ to $1$, in the $L_2$-edge sector, the positive two-peak structures remain robust despite the changes in the weight, which is transferred from the lower to upper peak. However, in the $L_3$ sector, the inclusion of CVEC dramatically alters the overall shape of the spectra. Positive spectra emerge in the higher energy sector of the $L_3$ edge for both dichroism, eventually evolving into the positive peak in the $L_3$ edge, which corresponds to the XMCD peak at around 644.6 eV in the experiments (See Fig.~\ref{Fig1}(a)). Also, the two negative peaks rearrange into one large negative peak with subpeak structures. Once CVEC is included, the XMCD spectra \emph{cannot} be simply interpreted as the dipole transition between occupied core and unoccupied valence electronic bands, and the conventional analysis on the XMCD sign, hence the sum-rule, is no longer valid. As we have individually analyzed, eventually, one need to fully include the CT and CVEC as well as JT distortion for the proper simulation the experimental XMCD spectra (Fig.~\ref{Fig_mcd_th}(b)).

 One can obtain the microscopic insight from the PED behaviors for the full inclusion of CT and CVEC effects ($r=1, s=1$) case. Figures~\ref{Fig_dst}(f) and (g) present the PED behaviors for $D_{\pm}^{t_{2g}}  \big| t_{2g\uparrow}^3\varphi_{x^2\uparrow}  \big>$ and $D_{\pm}^{e_g} \big| t_{2g\uparrow}^3 \varphi_{x^2\uparrow} \big>$, respectively. The CVEC arises from the atomic electron-electron correlations between core $2p$ and valence $3d$ orbitals, allowing exchanges between $p_{1/2}$ and $p_{3/2}$ core holes along with simultaneous exchanges among valence $d$ orbitals. This leads to a fine structure of the final multiplet states in the XAS process. Consequently, the PEDs are much extended, and the XMCD spectra exhibit complex behaviors.

 The dominant negative peak at around 3.4 eV and the positive one at 4.8 eV in the $L_3$ edge XMCD can be attributed to the dipole transition from $p_{3/2}$ to $e_g$ orbitals. And we see that the $L_2$-edge XMCD spectra near 14.3 eV are almost compensated by positive and negative contributions from the dipole transitions from $p_{1/2}$ to $t_{2g}$ and $e_g$ orbitals, respectively, and the positive peaks around 13.1 eV and 16 eV mainly originated from the dipole transition from $p_{1/2}$ to $t_{2g}$ and $e_{g}$ orbitals, respectively. These observations highlight the intricate effects of CVEC on the PED, hence, on the XMCD spectra, emphasizing the need for a detailed analysis of these interactions to understand the underlying electronic transitions.

 For the microscopic understanding of full CT effects, we investigate the PED behaviors for the dipole transition for two ligand-hole state $D_{\pm} \big| t_{2g\uparrow}^3e_{g\uparrow}^2L_{e_g\uparrow}^2\big>$ for Mn$^{4+}$, and one and two ligand-hole states $D_{\pm} \big| t_{2g\uparrow}^3 e_{g\uparrow}^2 L_{e_g\uparrow}^1 \big>$ and $D_{\pm} \big| t_{2g\uparrow}^3 e_{g\uparrow}^2 t_{2g\downarrow}^1 L_{e_g\uparrow}^1 L_{t_{2g}\downarrow}^1\big>$ for Mn$^{3+}$ (see Supplementary Fig. S7).
Unlike the cases for $\big| t_{2g\uparrow}^3 \big>$, $\big| t_{2g\uparrow}^3 e_{g\uparrow}^1 \big>$, and $\big| t_{2g\uparrow}^3 e_{g\uparrow}^1 L_{e_g\uparrow}\big>$, the dipole transition from $p_{3/2}$ and $p_{1/2}$ to $e_{g\uparrow}$ orbitals is prohibited because all $e_{g\uparrow}$ orbitals are fully occupied.
We clearly show the reduction of PED intensity of $D_{\pm}^{e_g} \big| t_{2g\uparrow}^3e_{g\uparrow}^2L_{e_g\uparrow}^2\big>$, $D_{\pm}^{e_g} \big| t_{2g\uparrow}^3 e_{g\uparrow}^2 L_{e_g\uparrow}^1 \big>$, and $D_{\pm}^{e_g} \big| t_{2g\uparrow}^3 e_{g\uparrow}^2 t_{2g\downarrow}^1 L_{e_g\uparrow}^1 L_{t_{2g}\downarrow}^1\big>$ states takes place at around 0.6 and 12 eV for the case of $r=1, s=0$.
Note that the negative (positive) contributions to the $L_3$-edge ($L_2$-edge) XMCD are generated by the dipole transition from $p_{3/2}$ ($p_{1/2}$) to $e_{g\uparrow}$ orbitals as shown in Fig.~\ref{Fig_dst}(d) and Supplementary Fig. S5(d)
Unless the CT are appropriately included, that contribution can be overestimated.
Therefore, we ascertain the importance of the full CT inclusion, which provides complete description of the XMCD behaviors -- notably the positive two XMCD peaks at the $L_2$ edge near 13.1 eV and 16 eV that the weighted-sum approach with sole $d^{n+1}L^1$ inclusion cannot achieve. Especially for the systems with significant mixed-valence fluctuations, our findings underscore the importance of full CT effects consideration for the accurate description of the XMCD spectra.

%===========================================
\begin{figure}[!t]
\centering
\includegraphics[width=0.95\columnwidth]{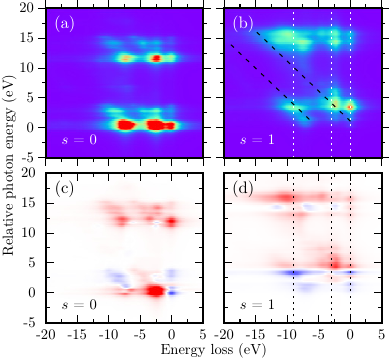}
\caption {
 Theoretical resonant inelastic X-ray scattering (RIXS) map for depending on the scale parameter ($s$) of the CVEC channel (a) $s=1, s=0$ and (b) $r=1, s=1$, and their MCD map for (c) $r=1, s=0$ and (d) $r=1, s=1$.
The vertical dotted lines in (b) and (d) indicate energy losses of 0, $-3$, and $-9$ eV. The black dotted lines denote the emission cuts shown in Fig.~\ref{Fig1}(c). The relative photon energy of 0 eV corresponds to 639.2 eV.
}
\label{Fig_rxs_th}
\end{figure}
%===========================================

\subsection{Theoretical RIXS-MCD spectra}

 Now we analyze the RIXS and RIXS-MCD spectra, where the importance of the CVEC is further emphasized. As in the experimental X-ray geometry setup as shown in Fig.~\ref{Fig1}(b), we calculated the spectra assuming the incident photon has the LCP or RCP. (see Supplementary Note 4 for the calculation details). In Figs.~\ref{Fig_rxs_th}(a)--(d), we present the calculated RIXS and RIXS-MCD maps as functions of incident photon energy and outgoing photon energy loss for the cases of the absence ($s=0$) and presence of the CVEC ($s=1$). We find that our calculation with full consideration of the CVEC in the strong $d$-$p$ hopping regime, shown in Fig.~\ref{Fig_rxs_th}(b) and (d), successfully reproduce the experimental RIXS and RIXS-XMCD in Fig.~\ref{Fig1}(c) and (d).

 In both $s=0$ and $1$ cases, the calculations yield identical multiplet structures for the initial and final states in the RIXS process, which shows that the CVEC is inert unless the core holes are populated. Consequently, the positions of RIXS and RIXS-MCD peaks with respect to the energy loss remain the same.

%===========================================
\begin{figure*}[!t]
\centering
\includegraphics[width=0.95\textwidth]{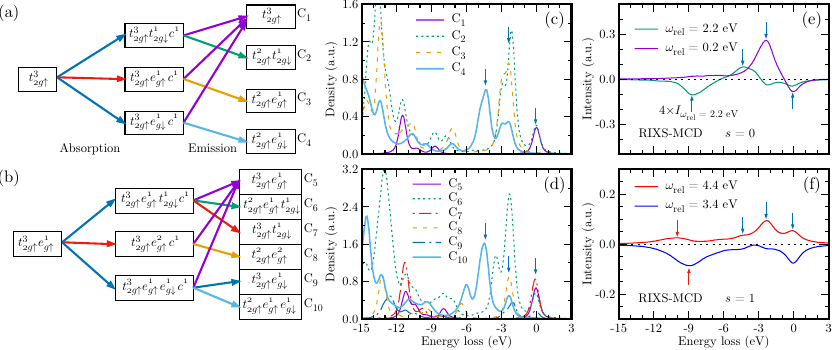}
\caption {
Schematic diagram of the resonant inelastic X-ray scattering (RIXS) process from relevant ground configurations: (a) $t_{2g\uparrow}^3$ and (b) $t_{2g\uparrow}^3e_{g\uparrow}^1$.
The partial excitation distribution (PED) behaviors of final states derived from initial (c) $\big| t_{2g\uparrow}^3 \big>$ and (d) $\big| t_{2g\uparrow}^3e_{g\uparrow}^1 \big>$ states.
C$_1$ -- C$_{10}$ represent the configurations of final states given in (a) and (b).
The theoretical RIXS-MCD spectra when the relative photon energies of incident X-ray are (e) $0.2$ and $2.2$ eV for $s=0$, and (f) $3.4$ and $4.4$ eV for $s=1$. These energies are consistent with the peak positions of the measured XMCD spectra.
Arrows in (c) -- (f) indicate the peak positions of RIXS-MCD spectra.
Other parameters are presented in Table~\ref{tb_para}.
}
\label{Fig_spes}
\end{figure*}
%==================

 However, the details of the peak structures, such as the sign and intensities, exhibit markedly different behaviors depending on the CVEC strengths. The CVEC strongly affect the intermediate states of the RIXS process, and split them into finer structures, thereby modifying the effective selection rule for a given photon energy. As a result, when the CVEC effect is included, the RIXS-MCD maps extend over wider ranges of the incident photon energy. Note that the negative $L_3$ edge at the elastic part shifts downward while the positive portion at the inelastic part moves upward (see Fig.~\ref{Fig_rxs_th}(c) and (d)).

 For a comprehensive analysis on the RIXS-MCD spectra, we explore the transitions of ground multiplet configurations during the absorption and emission processes in RIXS. Figures~\ref{Fig_spes}(a) and (b) illustrate the transitions starting from the initial configurations $t_{2g\uparrow}^3$ and  $t_{2g\uparrow}^3e_{g\uparrow}^1$. As a result, the final states derived from $\big| t_{2g\uparrow}^3 \big>$ are mainly attributed to the states with $t_{2g\uparrow}^3$, $t_{2g\uparrow}^2t_{2g\downarrow}^1$, $t_{2g\uparrow}^2e_{g\uparrow}^1$, and $t_{2g\uparrow}^2e_{g\downarrow}^1$ configurations. Similarly, the final states derived from $\big| t_{2g\uparrow}^3e_{g\uparrow}^1 \big>$ are mainly attributed to the states with $t_{2g\uparrow}^3e_{g\uparrow}^1$, $t_{2g\uparrow}^2e_{g\uparrow}^1t_{2g\downarrow}^1$, $t_{2g\uparrow}^3t_{2g\downarrow}^1$, $t_{2g\uparrow}^2e_{g\uparrow}^2$, $t_{2g\uparrow}^3e_{g\downarrow}^1$, and $t_{2g\uparrow}^2e_{g\uparrow}^1e_{g\downarrow}^1$  configurations. With these setups, we can calculate the microscopic PED behaviors of the relevant final states involved in the RIXS-MCD process as shown in Figs.~\ref{Fig_spes}(c) and (d).

 We observe that the local $dd$ excitations primarily contribute to the RIXS spectra for $-1$ -- $-5$ eV , while CT excitations become more prominent for $-6$ -- $-15$ eV, possibly due to enhanced Coulomb interactions. Additionally, the RIXS-MCD peaks above 1 eV for both $s=0$ and $1$ cases are mainly attributed to the excited states with $t_{2g\uparrow}^2e_{g\uparrow}^1$, $t_{2g\uparrow}^2e_{g\downarrow}^1$
 as well as $t_{2g\uparrow}^2e_{g\uparrow}^2$, and $t_{2g\uparrow}^2e_{g\uparrow}^1e_{g\downarrow}^1$ for the $dd$ excitations. As shown in Fig.~\ref{Fig_spes}(c) and (d), these correspond to transition from $t_{2g}$ to $e_g$ orbitals, with or without spin-flip transition. Furthermore, the CT excited states involving spin-flip transtions, characterized by configurations such that $t_{2g\uparrow}^2e_{g\uparrow}^1 t_{2g\downarrow}^1 L_{e_{g}\uparrow}^1$, $t_{2g\uparrow}^3 e_{g\downarrow}^1 L_{e_{g}\uparrow}^1$, $t_{2g\uparrow}^2 e_{g\uparrow}^2 t_{2g\downarrow}^1 L_{e_{g}\uparrow}^1$, and $t_{2g\uparrow}^3 e_{g\uparrow}^1 e_{g\downarrow}^1L_{e_{g}\downarrow}^1$, are heavily populated at thse enegy losses, giving main RIXS-MCD intensity (see Supplementary Fig. S8).
 The peak positions of their PED spectra are well matched with those of RIXS-MCD spectra, and we can clearly identify the origins of the multiple features.

 Here, we want to emphasize that the detailed behaviors show a strong dependency on the relative photon energy $\omega_{rel}$ and CVEC effects. For instance, the RIXS-MCD spectra without the CVEC effect, $s=0$, and for $\omega_{rel}=0.2$ eV exhibit positive peaks at around $-2.4$ eV of energy loss, while for $s=0$ and $\omega_{rel}=2.2$ eV they show negative peaks and additional positive peaks at around $-4.4$ eV of energy loss. If one includes the full CVEC effect, $s=1$, and for $\omega_{rel}=3.4$ eV, negative peaks emerge at about $-2.4$ and $-4.4$ eV in the RIXS-MCD spectra. In contrast, positive two peaks shown at approximately the same energy losses, which is consistent with the peak positions of PED of the $t_{2g\uparrow}^2e_{g\downarrow}^1$ ($t_{2g\uparrow}^2e_{g\uparrow}^1e_{g\downarrow}^1$) configuration for $s=1$ and $\omega_{rel}=4.4$ eV (see Fig.~\ref{Fig_spes}(e) and (f)).
 These observations directly support that CVEC effect predominantly determines the energies and symmetries of the intermediate states, which eventually structures the RIXS-MCD selections. Here, we clearly demonstrated our model can identify the details of the signals, and the inelastic component measurement via RIXS-MCD would be considerable to be an alternative way for exploring the microscopics of the complex systems.

\section{Conclusion}

 In conclusion, by employing AIM calculation that fully accounts for the CT and many-body CVEC as well as the JT distortion, we successfully reproduce the key features of XMCD and RIXS-MCD spectra for epitaxial LSMO films, a representative itinerant ferromagnet. Our analysis highlights the crucial roles of full CT treatment, particularly in the mixed valence systems, where the CT approaches based on restricted $d^{n+1}L^1$ CT configurations often fail.
 We also reveal that the JT effect of Mn$^{3+}$ ions is crucial to reproduce the XMCD spectra of metallic LSMO systems.
 Furthermore, we demonstrate that the CVEC effect significantly modulates the multiplet structures of the final states of XMCD and intermediate states of RIXS-MCD, leading to complex dipole selection behaviors dependent on incident X-ray photon energy. Note that as we have demonstrated here, CVEC-induced complexity greatly reduces the practicality of the XMCD sum-rule, particularly in light 3$d$ systems, and emphasizes the strong photon energy dependency of the RIXS-MCD map.

 In our approach, starting from the atomic model, we develop many-body calculations that include the CVEC effect. To incorporate the metallic feature of valence electrons, we employ full CT based on ligand field theory, offering a significant improvement over conventional treatments of mixed-valence systems. Our study represents an important step toward extracting accurate empirical physical parameters from spectral data, such as those from RIXS and XMCD. Moving forward, parameterizing AIM using density functional theory  and dynamical mean-field theory methods offers a promising route for further refining our model and achieving even more precise agreement with experimental spectra~\cite{Haverkort2016,Hariki2018,Wang2019,Hariki2020,Haverkort2012}. Amalgamating two distinct approaches facilitates a more accurate reproduction of the experimental spectra. We believe our work is a significant advancement toward this goal, bridging theoretical and experimental investigations in complex correlated systems.

\section{Methods}

\subsection{Experiments}

Epitaxial LSMO films with 20 unit cells thick are grown on (001) SrTiO$_3$ single-crystal substrates via pulsed laser deposition. The laser fluence is relatively low, approximately 0.3\,J/cm$^2$, at the target surface with a spot size of about 10\,mm$^2$, and a repetition rate was 2\,Hz. During the film deposition, the temperature is maintained at 800\,$^\circ$C with an oxygen partial pressure of $5 \times 10^{-6}$\,Torr. After deposition, the samples are cooled down to room temperature avoiding exposure to air. Detailed growth conditions can be found in a previous publication~\cite{song2008enhanced}. XMCD and RIXS-MCD measurements are conducted on the films at beamline 13-3, Stanford Synchrotron Radiation Lightsource (SSRL).

 The XMCD spectra are obtained in the total electron yield (TEY) mode. The X-ray absorption spectroscopy (XAS) spectra of LSMO are measured with two orthogonal polarizations: right circular polarization (RCP) and left circular polarization (LCP), under a constant magnetic field ($H\sim 0.2$ Tesla) (see the inset of Fig.~\ref{Fig1}(a)).

 Traditionally, the RIXS-MCD signal is weak due to its small cross-section~\cite{ament2011resonant}. To address this limitation, we employ a high-efficiency superconducting transition-edge-sensor (TES) spectrometer~\cite{doriese2017practical}. Figure~\ref{Fig1}(b) illustrates the experimental setup for RIXS-MCD measurement using the TES under an external magnetic field ($H \sim$ 0.2 Tesla). In the RIXS process, the circularly polarized X-ray photons with a photon energy of $E_{in}$ are injected into the sample at the incident angle of 30$^\circ$ against the surface. Core electrons in the Mn $2p$ levels are excited to unoccupied states of the valence Mn $3d$ bands, resulting in a $2p$ $\rightarrow$ $3d$ dipole transition. As the excited electrons decay, X-ray photons are emitted with a certain photon energy of $E_{\rm out}$. The inelastic decay process ($\Delta E = E_{\rm out}-E_{\rm in} \ne 0$) enables the probing of electronic excitations such as phonon, magnon, $dd$, and charge-transfer (CT) excitations~\cite{Kotani2001,ament2011resonant}.

\subsection{Theoretical calculation}

 To gain microscopic insights into the XMCD and RIXS-MCD spectra, we employ the CI calculation based on the AIM, which is importantly employed for strongly correlated materials~\cite{Gunnarsson1983,Gunnarsson1983_2,Kotani1988,Jo1988,deGroot2008,Lee2000,Kotani2012}.
 In the model, the CT effect of ligand orbitals are considered as a bath. Notably, the XAS and RIXS spectra are primarily governed by local electronic correlations involving the Mn core $2p$ and valence $3d$ orbitals, as well as strong $pd$ hybridization between Mn $d$ orbitals and ligand $p$ orbitals. In contrast, the ligand bandwidth is believed to play a more subtle, secondary role, as is typical in systems like LSMO. Analogous to conventional CI calculations~\cite{deGroot2005}, we incorporate five ligand (bath) orbitals with identical energy $e_L$, whose symmetric properties are the same as those of five $d$ orbitals. For the overlap integrals, we assume $\sigma$-type ($\pi$-type) overlaps between $3d$ and bath orbitals with the $e_g$ ($t_{2g}$) symmetry. In line with the CI approach, we assign overlap strength of $\sqrt{3}V_{pd\sigma}$ for $e_g$ orbitals and $2V_{pd\pi}$ for $t_{2g}$ orbitals, where $V_{pd\sigma}$ and $V_{pd\pi}$ denote the interatomic matrix elements for $\sigma$ and $\pi$ bonds, respectively.

%===============================
\begin{table*}[!bt]
\caption{
Physical parameters for the Anderson impurity model in the eV unit.
To simulate the C-type Jahn-Teller order, we set $\theta_{JT}=2\pi/3$.
}
\label{tb_para}
\begin{ruledtabular}
\begin{tabular}{c c c c c c c c c c c c c c c }
      & $10Dq$ & $\Delta_{JT}$ & $\lambda$ & $F_{dd}^2$ & $F_{dd}^4$ & $\lambda_c$ &
 $F_{pd}^2$ & $G_{pd}^1$ & $G_{pd}^3$ & $\Delta$ & $U_{dd}$ & $U_{pd}$ &
 $V_{pd\sigma}$ & $V_{pd\pi}$ \\
 \hline
 Mn$^{4+}$ & $1.4$ & $0.1$ & $0.052$ & $9.933$ & $6.256$ & $7.500$ & $6.126$ & $4.621$ & $2.630$ & $1.5$ & $5$ & $6$ & $-2$ & $1$ \\
 Mn$^{3+}$ & $1.4$ & $0.1$ & $0.046$ & $9.132$ & $5.718$ & $7.500$ & $5.590$ & $4.143$ & $2.356$ & $1.5$ & $5$ & $6$ & $-2$ & $1$
\end{tabular}
\end{ruledtabular}
\end{table*}
%===============================

 Following is the full AIM Hamiltonian we consider:
\begin{equation}
 H = H_d + H_c + H_{dc} + H_L + H_t,
\end{equation}
where \\
\begin{tabular}{ll}
\begin{minipage}[t]{0.08\columnwidth}
$H_d$ :
\end{minipage} &
\begin{minipage}[t]{0.85\columnwidth}
Hamiltonian of Mn $3d$ valence orbitals characterized by the cubic crystal field ($10Dq$), Jahn-Teller distortion ($\Delta_{JT}$, $\theta_{JT}$), spin-orbit coupling ($\lambda$), long-range magnetic field $h$, and Coulomb interactions ($U_{dd}$, $F_{dd}^2$, $F_{dd}^4$).
\end{minipage} \\
\begin{minipage}[t]{0.08\columnwidth}
$H_c$ :
\end{minipage} &
\begin{minipage}[t]{0.85\columnwidth}
Hamiltonian of Mn $2p$ core orbitals characterized by
the spin-orbit coupling ($\lambda_c$) and energy level ($e_c$).
\end{minipage} \\
\begin{minipage}[t]{0.09\columnwidth}
$H_{dc}$ :
\end{minipage} &
\begin{minipage}[t]{0.85\columnwidth}
Hamiltonian of core-valence exchange correlations characterized by Slater-Condon parameters ($U_{pd}$, $F_{pd}^2$, $G_{pd}^1$, $G_{pd}^3$).
\end{minipage} \\
\begin{minipage}[t]{0.09\columnwidth}
$H_L$ :
\end{minipage} &
\begin{minipage}[t]{0.85\columnwidth}
Local energy of ligand (bath) orbital characterized by the energy level $\epsilon_L$.
\end{minipage} \\
\begin{minipage}[t]{0.09\columnwidth}
$H_t$ :
\end{minipage} &
\begin{minipage}[t]{0.85\columnwidth}
Hopping Hamiltonian between valence $3d$ and ligand (bath) orbitals characterized by $V_{pd\sigma}$ and $V_{pd\pi}$.
\end{minipage}
\end{tabular}

 Slater-Condon parameters ($F_{dd}^2$, $F_{dd}^4$, $F_{pd}^2$, $G_{pd}^1$, $G_{pd}^3$) are typically set at 80\% of Hatree-Fock values (atomic values)~\cite{stavitski2010ctm4xas}.
 The Jahn-Teller (JT) distortion is described by two parameter for tetragonal distortions, i.e., $\Delta_{JT}$, and $\theta_{JT}$. The level splitting of the $e_g$ orbitals is determined by $\Delta_{JT}$.
 The orbital angle $\theta_{JT}$ is taken to be 0, $2\pi/3$, and $4\pi/3$, corresponding to  the elongation of the octahedron along the $z$, $x$, and $y$ axes, respectively. Accordingly, $z^2$ and $x^2-y^2$ orbitals, $x^2$ and $y^2-z^2$ orbitals, and $y^2$ and $z^2-y^2$ orbitals are split such that
 $\Delta_{JT}=E_{z^2}-E_{x^2-y^2}$, $\Delta_{JT}=E_{x^2}-E_{y^2-z^2}$, and $\Delta_{JT}=E_{y^2}-E_{z^2-x^2}$, respectively. Since LaMnO$_3$ exhibits the C-type JT distortiopn, $\theta_{JT}$ is taken to be $2\pi/3$ or $4\pi/3$ (both cases give the same results).

 The spin-orbit coupling parameter of Mn core orbitals $\lambda_c$ is selected for fitting the experimental XAS splitting of $L_2$ and $L_3$ edges optimally.
 The Coulomb repulsion parameters $U_{dd}$ and $U_{pd}$ are defined as $U_{dd} = F_{dd}^0 - \frac{2}{63}(F_{dd}^2+F_{dd}^4)$ and $U_{pd} = F_{pd}^0 - \frac{1}{15} G_{pd}^1 - \frac{3}{70} G_{pd}^3$, respectively~\cite{boca1999theoretical}.
 As a common approximation, We set $U_{pd}$ to be $1$ eV higher than $U_{dd}$~\cite{deGroot2005}.
 The CT energy $\Delta$ in $d^{n}$ atomic multiplets is defined as the difference in average energy between the $d^n$ and $d^{n+1}{L}^1$ configurations, where ${L}$ refers to the ligand (bath) hole orbitals. This is obtained as $\Delta = E(d^{n+1}{L}^1) - E(d^n) = n U_{dd}+6U_{pd} - \epsilon_L$, assuming all ligand (bath) hole orbitals have the same energy of $\epsilon_L$.

 In the calculation, $\epsilon_L$ is automatically set according to $\Delta$, $U_{dd}$, and $U_{pd}$ values.
To incorporate the magnetic ordering effect of LSMO, we apply an auxiliary magnetic field of $\mu_B h = 0.01$ eV, which is much smaller than the ordering temperature of the LSMO and corresponds to the mean-field from neighboring ferromagnetic spins. The remaining parameters are adjusted to fit the experimental spectra. The specific parameters used are presented in Table~\ref{tb_para}.

Our choice of $10Dq = 1.4$ eV is consistent with conventional values ranging from $1$ to $1.5$ eV~\cite{yu2010interface,Shibata2018,Kuepper2012,Hariki2016}. We adopted a CT energy of $\Delta = 1.5$ eV. That is lower than the values previously reported in some literature for LSMO films ($\Delta = 4 \sim 5$ eV)~\cite{yu2010interface,Shibata2018,Hariki2016} while it is larger than the other selection of negative value for. However, as demonstrated in Supplementary Fig.~S5, the relative intensity ratio of the positive $L_2$ peaks in the XMCD spectra is highly sensitive to this parameter. Our selected value of $\Delta = 1.5$ eV yields a more consistent agreement with the experimentally observed intensity ratio for given parameter set.
Our selected values of $U_{dd}$ and $U_{pd}$ are slightly smaller than the values of $U_{dd}=5.5$ and $U_{pd} = 6.5$ eV in Ref.~[\onlinecite{Hariki2016}], while the strength of $V_{pd\sigma}$ is slightly larger in magnitude than the value of $-1.7 \sim -1.8$ eV.

As we apply the CI-based methodology to the metallic LSMO system, the treatment of the mixed-valence nature of Mn ions--and consequently, an improved description of the CT effect—is essential.
In our model, we consider the CT effect from the ligand (bath) states.
Given the atomic-like nature of the core orbitals and the extremely
short timescale of the resonant process, the intermediate and final states in XAS and RIXS processes are predominantly sensitive to the local electronic configurations at the Mn absorbent site.
We conduct independent calculations for the Mn$^{4+}$ and Mn$^{3+}$ by starting with the initial $d^3L^0$ and $d^4L^0$ configurations and subsequently including all higher-order CT states up to $d^{10} L^{7}$ and $d^{10}L^{6}$, respectively.
The overall spectra for the metallic LSMO system are then reproduced via the weighted sum of these two calculations.
 The simplest treatment for CT effects is to restrict the CT space to only $d^{n+1}L^1$ configurations.
 While this method is computationally inexpensive and efficient for systems with weak CT effect, the CT treatments beyond simple inclusion with $d^{n+1}L^1$ configurations has been demonstrated to be essential to interpret accurately the ground multiplet states of various mixed-valence systems and its core-valence spectroscopy~\cite{Arenholz2006,Wang2022,Hariki2016,Hariki2022,Takegami2025}.
 Therefore, our approach, which incorporates a full CT consideration, delivers a critically improved description of the experimental spectra.

The AIM is solved by applying the exact diagonalization based on the Lanczos method~\cite{Wu2000}, and we calculate the ground state and its energy.
Based on our ground state, we calculate the XAS and RIXS spectra. The details on XAS and RIXS calculations are described in Supplementary Note 3 and 4, respectively.
The physical parameters, listed in Table~\ref{tb_para}, are optimized to ensure
that the calculated spectra closely resemble the experimental ones (see Supplementary Note 5, 6, and 7).
Furthermore, our calculation of the $d$-band electronic structures successfully
captures the dominant peak structures of those from the density functional theory calculation (see  Supplementary Note 2 for the details).
These findings demonstrate that our AIM describes the essential electronic features and confirm that our CI calculation effectively captures the mixed-valence fluctuations in LSMO systems.

\bibliography{references}

\section{Acknowledgments}
We thank K.-T. Ko, Hyeong-Do Kim, and Atsushi Hariki for the interesting discussions. B.H.K and B.K. thank the Center for Theoretical Physics of Complex Systems (IBS-PCS) Advanced Study Group program for their support during this collaboration.
All soft x-ray experiments were carried out at the SSRL (beamline 13-3), SLAC National Accelerator Laboratory. This study at the SSRL/SLAC is supported by the U.S. Department of Energy, Office of Science, Office of Basic Energy Sciences under contract no. DE-AC02-76SF00515. B.H.K was supported from the Institute for Basic Science in the Republic of Korea through the Project No. IBS-R024-D1. B.K. acknowledges support from NRF (Grants No. RS-2023-00256050, No. RS-2024-00401881, No. RS-2022-NR068223, and No. RS-2026-25472078) and KISTI Supercomputing Center (Project No. KSC-2023-CRE-0413).
HH acknowledges support from the Science and Technology Commission of
Shanghai Municipality Grant No. 21JC1400200. PA acknowledges support from the Gordon and Betty Moore Foundation’s EPiQS Initiative, Grant GBMF9452.
The work at SNU was supported by the Leading Researcher Program of the National Research Foundation of Korea (Grant No. RS-2020-NR049405).
We thank the Center for Advanced Computation in KIAS for providing computing resources.

\section{Data availability}
The data that support the findings of this study are provided in the main text. The raw data is available from the corresponding author upon request.

\section*{Author contributions}
B.H.K., S.-J.L., J.-S.L., and B.K. conceived the research project. B.H.K. and B.K. developed the theoretical framework; S.-J.L. and J.-S.L. performed the experiments. B.H.K, S.-J.L, J.-S.L and B.K. discussed the results and contributed to writing the manuscript, and all approved the final version

%=========================

\clearpage
\onecolumngrid
%\newpage

\section*{Supplementary Information:\\
X-ray magnetic circular dichroism and resonant inelastic X-ray scattering explained:\\
role of many-body correlation and valence fluctuations}

\renewcommand{\figurename}{{\bf Supplementary Figure}}
\renewcommand{\thetable}{S\arabic{table}}
\renewcommand{\thefigure}{S\arabic{figure}}
\renewcommand{\theequation}{S\arabic{equation}}
\setcounter{table}{0}
\setcounter{figure}{0}
\setcounter{equation}{0}
\renewcommand{\bibnumfmt}[1]{[S#1]}
\renewcommand{\citenumfont}[1]{S#1}
\def\bibsection{\section*{Supplementary References}}
\renewcommand{\thesection}{Supplementary Note \arabic{section}}

\makeatletter
\def\@hangfrom@section#1#2#3{\@hangfrom{#1#2}#3} %\MakeTextUppercase{#3}}
\def\@hangfroms@section#1#2{#1#2} %\MakeTextUppercase{#2}}
\makeatother

\subsection{Ground charge-transfer configurations}
\label{appen:AIM}

%===============================
\begin{figure}[!b]
\centering
\includegraphics[width=0.6\columnwidth]{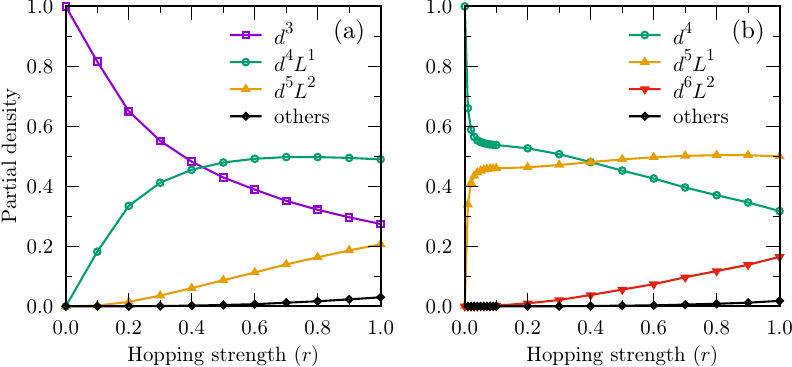}
\caption {
% (a) The partial density of states with various electronic configurations
%for the ground state and (b) average number of holes at the Mn site
The partial density of states with various electronic configurations
for the ground state of the Anderson impurity model for (a) Mn$^{4+}$ and (b) Mn$^{3+}$
as a function of the scale parameter $r$ of hopping strength.
$r$ is defined as $V_{pd\sigma}=-2 \times r$ and $V_{pd\pi} = 1 \times r$ eV.
}
\label{Fig_dn}
\end{figure}
%===============================

% Figure~\ref{Fig_dn} presents the partial density of states with various electronic configurations and average number of holes at the Mn site for the ground state, where we have systematically changed the hopping strengths between $3d$ and ligand orbitals.
% Here, we have defined the scale parameter $r$ which tunes the hopping strengths:
% $V_{pd\sigma}=-2 \times r$ and $V_{pd\pi} = 1 \times r$ eV.
% For example, in the case of $r=0$, there is no $d$-$p$ overlap, and the ground state is solely determined by states with $d^3$ configuration. Hence, the average number of holes is 7.
% When $r$ increases starting from $r=0$, electrons can dynamically transfer from ligand (bath) to Mn sites.
% The contribution from other states besides $d^3$ configurations contributes to the ground state and the number of holes decreases monotonically.
% Upon increasing the hopping strengths, more electrons move from the ligand to Mn sites, and when $r$ is about 0.6, transferred electron numbers reaches to $0.7$.
% Then $d^4{L}^1$ configuration contributes more to the ground state than $d^3$ configuration.
% The actual LSMO case corresponds to the $r > 0.6$ in the calculation.
% In the calculation, we set $r=1$. In this case, the averaged number of holes at the Mn site in the ground state is calculated to be about 6.1, slightly deviating from the nominal value of 6.3. We validate our calculations by comparing the electronic structures of the AIM with those from mean-field density functional theory calculation as in the next section.

Figure~\ref{Fig_dn} presents the partial density of states at the Mn site, decomposed by electronic configuration, for the ground state of the Anderson impurity model. To systematically investigate the effect of $p$-$d$ hybridization for nominal Mn$^{4+}$ and Mn$^{3+}$ configurations, we introduce a scale parameter $r$ that tunes the hopping strengths such that $V_{pd\sigma}=-2 \times r$ and $V_{pd\pi} = 1 \times r$ eV.
For instance, in the case of $r=0$, the $d$-$p$ hopping vanishes, and the ground states are solely determined by $d^3$ and $d^4$ configurations for Mn$^{4+}$ and Mn$^{3+}$, respectively.
As $r$ increases starting from $r=0$, electrons can dynamically transfer from ligand (bath) to Mn sites, introducing contributes from higher-order charge-transfer (CT) configurations.
Consequently, the ground multiplet state becomes an admixture of $d^n L^0$, $d^{n+1} L^1$, and $d^{n+2} L^2$ configurations, reflecting the mixed-valence nature of the system. Specifically, as shown in Fig.~\ref{Fig_dn}, the calculated ground-state wavefunctions are composed of approximately 27\% $d^3 L^0$, 49\% $d^4 L^1$, 21\% $d^5 L^2$, and 3\% $d^6 L^3$ configurations for nominal Mn$^{4+}$, and 32\% $d^4 L^0$, 50\% $d^5 L^1$, 16\% $d^6 L^2$, and 2\% $d^7 L^3$ configurations for nominal Mn$^{3+}$.

\subsection{Electronic structure}
\label{appen:DOS}

\begin{figure}[!b]
\centering
\includegraphics[width=0.5\columnwidth]{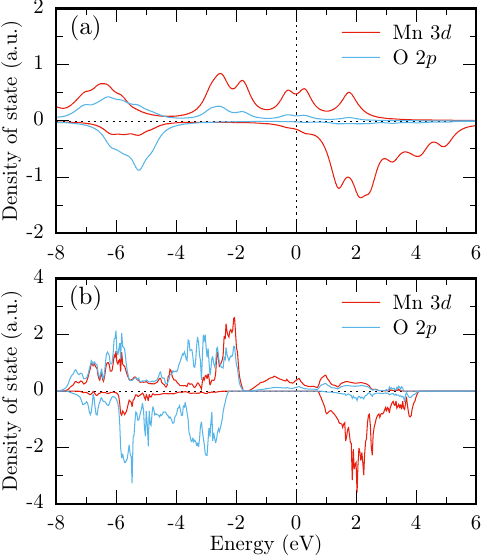}
\caption {
 The partial density of states calculated by (a) the Anderson impurity model (AIM) calculation and (b) the density functional theory calculation with the SCAN functional.
 The Fermi level is denoted by the vertical dotted lines. Parameters for the AIM are the same in the main manuscript.
}
\label{Fig_dos}
\end{figure}

To check the validity of our AIM calculation, we explore the electronic structures and compare them with those from the mean-field density functional theory (DFT) calculation.
To obtain the density of states (DOSs) from AIM, we first calculate the one-particle Green function
for a given frequency point $z$, which is expressed as follows:
\begin{equation}
G_{\alpha\beta} (z) =
   \left< \Psi_g \right| d_{\alpha} \frac{1}{z-H+E_g-\mu_c} d_{\beta}^\dagger
   \left| \Psi_g \right> +
   \left< \Psi_g \right| d_{\beta}^\dagger \frac{1}{z+H-E_g-\mu_c} d_{\alpha}
   \left| \Psi_g \right>,
\end{equation}
where $d_{\alpha}$ ($d_{\alpha}^\dagger$) is the annihilation (creation) operator
of valence $\alpha$ states and $\left| \Psi_g \right>$, $E_g$, and $\mu_c$ are
the ground state, ground energy, and chemical potential, respectively.
$\mu_c$ is determined from the Fermi level, which is to be located between electron and hole bands.
The partial DOS of $\alpha$ state can be computed by $A_\alpha(\omega) = -\frac{1}{\pi}\textrm{Im} G_{\alpha\alpha}(\omega+i\delta_b)$, where $\delta_b$ being the broadening parameter. We set $\delta_b$ to be 0.3 eV.
To simulate the spectra of La$_{0.7}$Sr$_{0.3}$MnO$_3$, we use a weighted sum of separate calculations for the Mn$^{4+}$ and Mn$^{3+}$ ions.
The DFT calculations are performed employing the Vienna ab initio simulation package (VASP).
We utilize SCAN (strongly constrained and appropriately normed semilocal) functional employing the realistic supercell of La$_{3/4}$Sr$_{1/4}$MnO$_3$ with eight formula units.
The energy cut for the plane waves is 400 eV and Monkhorst-Pack $k$-mesh of 4 $\times$ 4 $\times$ 4 was used.

Figure~\ref{Fig_dos} present the partial DOS calculated from the AIM and DFT.
We can clearly see that our AIM result qualitatively captures the main features of the DFT calculations:
i) the bonding bands of Mn-$3d$ and O-$2p$ orbitals are spread over $-7$--$-5$ eV, ii) spin-up antibonding bands are located at around $-2$ eV, and iii) the unoccupied spin-down bands of Mn-$3d$ orbitals have main peak at around $2$ eV with subpeak at $4$ eV.
Even though some details such as metallic bands at Fermi level and non-bonding O-$2p$ bands at around $-5$ -- $-2$ eV are different due to finite-size cluster limits, the essential features are well-captured from our AIM.
And we believe that the AIM calculation is capable to simulate the excitation spectra of LSMO with large energy scale ranges of about $10$ eV.

\subsection{XAS calculation}
\label{appen:XAS}

Let $\left| \Psi_g \right>$ be a ground state with energy $E_g$.
The XAS intensity for the circularly polarized X-ray photon $\mu_{m}$ ($m=\pm 1,0$) is calculated with the following equation:
\begin{equation}
\mu_{m} = -\frac{1}{\pi} \textrm{Im}
\big< \Psi_{g} \big| \hat{e}_{-m} \cdot \mathbf{D}^\dagger
\frac{1}{E_{\rm in}-H+ E_g + i\delta_{in}} \hat{e}_m \cdot \mathbf{D}
\big| \Psi_{g}\big>,
\label{Eq:XAS}
\end{equation}
where $E_{\rm in}$ is energy of incident x-ray with broadening $\delta_{in}$.
$\hat{e}_m$ is the x-ray polarization vector defined as
$\hat{e}_{\pm 1}= \frac{1}{\sqrt{2}}\left(\hat{x} \pm i \hat{y} \right)$ and $\hat{e}_{0}=\hat{z}$,
and $\mathbf{D}$ is the dipole operator given as
\begin{equation}
\mathbf{D} =  \sum_{\alpha p \sigma}
\left< \psi_{\alpha}^d \right| \mathbf{r} \left|\psi_m^c \right>
d^\dagger_{\alpha\sigma} c_{p \sigma},
\end{equation}
where $d^\dagger_{\alpha\sigma}$ and $d_{\alpha\sigma}$ ($c^\dagger_{p\sigma}$ and $c_{p\sigma}$) are the creation and annihilation
operators of valence $3d$ (core $2p$) orbitals with a wave function
$\psi_{\alpha}^d$ ($\psi_p^c$) and spin $\sigma$.
Then, XAS and XMCD intensities are  obtained from $\mu = \mu_{+1}+\mu_{-1}$ and $\Delta \mu =\mu_{+1}-\mu_{-1}$, respectively.

\subsection{RIXS calculation}
\label{appen:RIXS}

The RIXS spectrum is derived using the Kramers-Heisenberg equation:
\begin{equation}
I_{\rm RIXS}\left( \omega ,\mathbf{q},\bm{\epsilon},\bm{\epsilon}' \right) =
 \sum_f \left| \mathcal{F}_{fg}(E_{\rm in},\mathbf{q},
 \bm{\epsilon},\bm{\epsilon}^{\prime}) \right|^2
  \delta\left(\omega -E_f+E_g\right),
\end{equation}
where $\omega =E_{\rm in}-E_{\rm out}=-\Delta E $, $\mathbf{q}=\mathbf{k}'-\mathbf{k}$, and $E_g$ and $E_f$ are energies of ground and excitation states, respectively~\cite{ament2011resonant}.
Here, $\mathbf{k}$ ($\mathbf{k}'$), $E_{\rm in}$ ($E_{\rm out}$), and $\bm{\epsilon}$ ($\bm{\epsilon}'$) refer to the momentum, energy, and polarization of an incident (outgoing) X-ray photon, respectively.
Under the dipole approximation, the function $\mathcal{F}_{fg}$ is expressed as:
\begin{equation}
\mathcal{F}_{fg} \left( E_{\rm in},\mathbf{q},\bm{\epsilon},\bm{\epsilon}' \right) = \sum_{j} e^{i\mathbf{q}\cdot \mathbf{r}_j} \times
\big< \Psi_f \big| \bm{\epsilon}'^*\cdot \mathbf{D}_j^\dagger
\frac{1}{E_g+E_{\rm in} - H + i\delta_{\rm in}}
\bm{\epsilon} \cdot \mathbf{D}_j \big| \Psi_g \big>,
\end{equation}
where $\mathbf{r}_j$ and $\mathbf{D}_j$ are the position vector and dipole operator of the $j$th Mn ion, respectively. $\delta_{\rm in}$ is the broadening parameter of an incident X-ray photon.
The state $\left| \Psi \left( E_{\rm in},\mathbf{q},\bm{\epsilon},\bm{\epsilon}'\right) \right>$ can be expressed as:
\begin{equation}
\left| \Psi \left( E_{\rm in},\mathbf{q},\bm{\epsilon},
\bm{\epsilon}'\right) \right> =  \sum_{j} e^{i\mathbf{q}\cdot \mathbf{r}_j} \times
\bm{\epsilon}'^*\cdot \mathbf{D}_j^\dagger
\frac{1}{E_g+E_{\rm in} - H + i\delta_{\rm in}}
\bm{\epsilon} \cdot \mathbf{D}_j \left| \Psi_g \right>.
\label{Eq:Pin}
\end{equation}
Now, the RIXS intensity, $I_{\rm RIXS} \left(\omega,\mathbf{q},\bm{\epsilon},\bm{\epsilon}'\right)$, can be calculated from:
\begin{equation}
I_{\rm RIXS} \left(\omega,\mathbf{q},\bm{\epsilon},\bm{\epsilon}'\right) =
-\frac{1}{\pi} \textrm{Im} \big< \Psi \left( E_{\rm in},\mathbf{q},\bm{\epsilon},\bm{\epsilon}'\right) \big|
\frac{1}{\omega-H+ E_g + i\delta_b}
\big| \Psi \left( E_{\rm in},\mathbf{q},\bm{\epsilon},\bm{\epsilon}'\right) \big>,
\label{Eq:RIXS}
\end{equation}
where $\delta_b$ is the broadening parameter for the RIXS spectra.
Given that the experimental resolution of the RIXS measurement using a transition-edge-sensor spectrometer is about 1.5 eV, we set $\delta_b$ to be 0.75 eV. Additionally, we assume $\delta_{\rm in} = 0.3$ eV.
We utilize the Lanczos method to solve the RIXS intensity in Eq.~\ref{Eq:RIXS},
and the complex-shifted minimal residual method to solve for the state in Eq.~\ref{Eq:Pin}.

In the AIM calculation, we set $\mathbf{q}=0$ and obtain the RIXS intensity
for the circularly polarized incident X-raya photon ($m=\pm1$) as follows:
\begin{equation}
I_{\rm RIXS}^{m} \left( \omega \right) =
I_{\rm RIXS} \left(\omega, \bm{\epsilon}_{m}, \bm{\epsilon}'_{+}\right)
+I_{\rm RIXS} \left(\omega, \bm{\epsilon}_{m}, \bm{\epsilon}'_{-}\right)
\end{equation}
in the experimental geometry as shown in Fig.~1(b) in the main text.
The RIXS and RIXS-MCD intensities are calculated as $I_{\rm RIXS} = I_{\rm RIXS}^{+} + I_{\rm RIXS}^{-}$ and $I_{\rm RIXS-MCD} = I_{\rm RIXS}^{+} - I_{\rm RIXS}^{-}$, respectively.
Note that while the experimental momentum transfer deviates from $\mathbf{q}=0$, the elastic contribution in resonant X-ray scattering is significantly suppressed due to the magnetic structure factor and atomic form factor. Since the CI calculation is limited to $\mathbf{q}=0$ and cannot fully capture finite-momentum effects, the theoretical elastic peak is overestimated compared to the experiment. To account for this discrepancy, we omit the elastic contribution to simulate the RIXS and RIXS-MCD spectra.

% \section{Weighted sum from cluster calculation}
% \label{appen:CTM4XAS}
\subsection{Charge transfer effects}
\label{appen:CT_effect}

\begin{figure}[!htb]
\centering
\includegraphics[width=0.6\columnwidth]{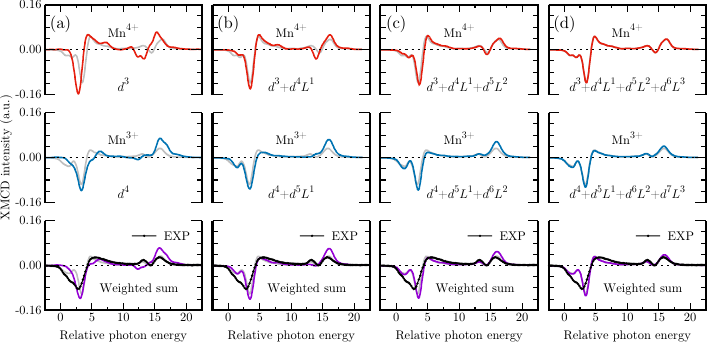}
\caption{
%The X-ray magnetic circular dichroism (XMCD) intensities of
%Mn$^{4+}$, Mn$^{3+}$, and their weighted sum (30\% contribution of Mn$^{4+}$ and 70\% contribution of Mn$^{3+}$) are presented
%with (a) no charge transfer (CT) states and (b) CT states, i.e.,
%$d^4L^1$ states for Mn$^{4+}$ and $d^5L^1$ states for Mn$^{3+}$.
%All results are obtained by the CTM4XAS software~\cite{stavitski2010ctm4xas}.
The X-ray magnetic circular dichroism (XMCD) intensities of
Mn$^{4+}$, Mn$^{3+}$, and their weighted sum (30\% contribution of Mn$^{4+}$ and 70\% contribution of Mn$^{3+}$) calculated (a) without the charge transfer (CT) state and CT configurations up to (b) $d^{n+1}L^1$, (c) $d^{n+2}L^2$, and $d^{n+3}L^3$. Here, $L$ density a ligand hole state.
The solid gray line represents the calculated XMCD spectra considering the full CT states from $d^nL^0$ to $d^{10}L^{10-n}$ configurations, and black circles corresponds to the experimental XMCD spectra.
}
\label{SFig_ctl}
\end{figure}

% To compare our XMCD results with {\cblue incoherent weighted-sum} approaches for mixed-valence states,
% we perform calculations using the CTM4XAS software~\cite{stavitski2010ctm4xas}.
% First, we calculate the XMCD spectra for Mn$^{4+}$ with a $d^3$ configuration and Mn$^{3+}$ with a $d^4$ configuration separately.
% We then take a weighted sum of the spectra, with 30\% contribution from Mn$^{4+}$ and 70\% from Mn$^{3+}$,
% representing the mixed-valence state of La$_{0.7}$Sr$_{0.3}$MnO$_3$.
% As shown in Fig.~\ref{SFig_ctm4xas_mcd}(a), this approach fails to reproduce
% the extended subpeak structures in the $L_3$-edge at the relative photon energy of around $0 \sim 2$ eV and the positive XMCD signal in the $L_2$ edge at $13 \sim 14$ eV regime.
% Next, we incorporate the CT effect into the calculation by adding the CT states: $d^{4}L^1$ for Mn$^{4+}$ and $d^{5}L^1$ for Mn$^{3+}$.
% Similarly, we take the weighted sum of the resulting spectra.
% These results, presented in Fig.~\ref{SFig_ctm4xas_mcd}(b), also fail to reproduce the extended subpeak structures at the $L_3$-edge and the distinctive positive two-peak structure at the $L_2$-edge observed in the experimental XMCD spectra.
% Therefore, the {\cblue weighted-sum} approaches, even with the inclusion of the CT effect of $d^{n+1}L^1$ configuration, are inadequate in capturing the consistent XMCD features observed experimentally.
To identify the role of CT effects more comprehensively in the XMCD spectra, we explore the spectral variation with respect to the included level of CT states. As shown in Fig.~\ref{SFig_ctl}, we observe that the CT states up to the $d^{n+3}L^3$ configuration must be included to achieve a spectral shape similar to that obtained from the full CT calculation.
Thus, weighted-sum approaches, even those including the $d^{n+1}L^1$ CT configuration, remain inadequate for capturing the consistent XMCD features observed experimentally.

%===========================================
\begin{figure}[!htb]
\centering
\includegraphics[width=0.9\columnwidth]{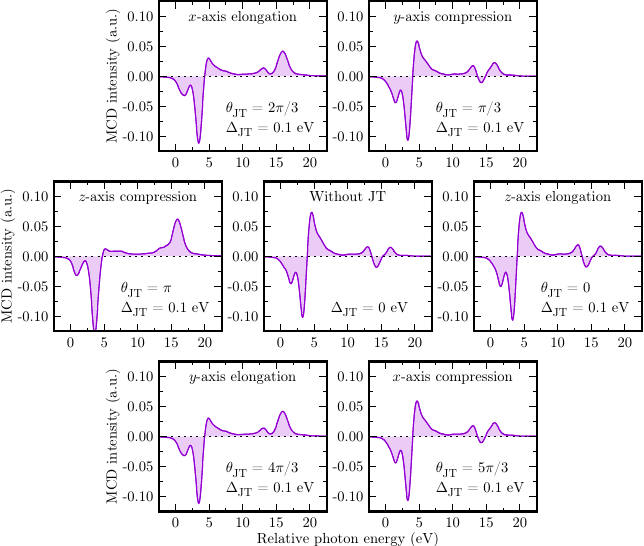}
\caption{
Calculated X-ray magnetic circular dichroism (XMCD) spectra for various Jahn-Teller parameters, $\Delta_{JT}$ and $\theta_{JT}$.
The center panel presents the result without JT distortion ($\Delta_{JT} = 0$ eV). The remaining panels show the XMCD spectra for various $\theta_{JT}$ values ($0$, $\pi/3$, $2\pi/3$, $\pi$, $4\pi/3$, and $5\pi/3$) at a fixed $\Delta_{JT} = 0.1$ eV.
The XMCD spectra is obtained through the weighted sum of separate calculations for Mn$^{4+}$ and Mn$^{3+}$ cases.
All other parameters are listed in Table I in the main text.
% The weighted sum of XMCD data in Fig.~\ref{SFig_ctm4xas_mcd} with the addition of JT effect. We set the splitting of $e_g$ orbitals to be (a) $E_{x^2-y^2}-E_{z^2} = 0.1$ eV for an octahedral elongation along the $z$ axis and (b) $E_{y^2-z^2}-E_{x^2} = 0.1$ eV for the elongation along the $x$ axis, respectively. Because results for the elongation along the $y$ axis are almost the same as those for the $x$ axis, they are not presented.
}
\label{WS_JT}
\end{figure}
%===========================================

\subsection{Jahn-Teller distortion}

The Jahn-Teller (JT) distortion plays an essential role in the electronic properties of magnetic systems containing Mn$^{3+}$ ions. Therefore, we investigate whether the characteristic two-peak structure at the $L_2$ edge could be attributed to the JT effect.
The JT distortion is parameterized by $\Delta_{JT}$ and $\theta_{JT}$.
Figure~\ref{WS_JT} presents the calculated XMCD spectra for various JT parameters.
Our results indicate that the JT distortion significantly influences the XMCD lineshape, particularly the positive two-peak structure at the $L_2$ edge.
Furthermore, the theoretical spectra for $\theta_{JT}=2\pi/3$ and $4\pi/3$ are consistent with the experimental observations, whereas those calculated for other $\theta_{JT}$ values deviates evidently. Note that the JT distortion for $\theta_{JT}=2\pi/3$ and $4\pi/3$ correspond to the elongation of MnO$_6$ octahedron along the $x$ and $y$ axes, respectively. This is consistent with the cooperative JT distortion observed in LaMnO$_3$ systems.
% In Fig.~\ref{WS_JT}, we have included the JT distortion effect in the weighted sum, where we still find that there are major discrepancies. The characteristic positive two-peak structure at the $L_2$-edge is not reproduced as before.

\subsection{XMCD spectra for various parameters}

\begin{figure}[!htb]
\centering
\includegraphics[width=1.0\columnwidth]{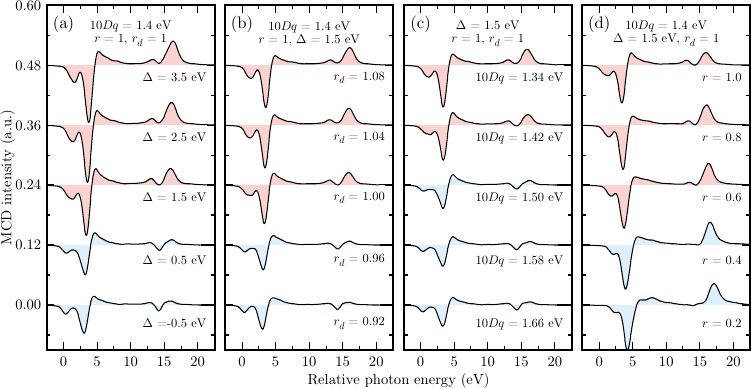}
\caption{
Calculated X-ray magnetic circular dichroism (XMCD) spectra as a function of
for various physical parameters: (a) the charge-transfer energy $\Delta$, (b) the Slater-Condon scale parameter $r_d$, (c) the cubic crystal-field strength $10Dq$, and (d) the hopping scale parameter $r$.
The Slater-Condon parameters $F_{dd}^2$ and $F_{dd}^4$ are obtained by multiplying $r_d$ with the reference values listed in Table I of the main text. The hopping strengths are defined by the scale parameter $r$ such that $V_{pd\sigma}=-2 \times r$ and $V_{pd\pi}= 1 \times r$. All other parameters are adopted in Table I of the main text.
}
\label{SFig_SP}
\end{figure}

Figure~\ref{SFig_SP} presents the calculated XMCD spectra as a function of various physical parameters, including the CT energy, cubic crystal-field strength, Slater-Condon parameters, and hopping strengths. We find that the theoretical spectra reproduce the experimental observations well when the CT energy is $\Delta = 1.5$ eV, the Slater-Condon parameters $F_{dd}^2$ and $F_{dd}^4$ are set to their atomic values (80\% of the Hartree-Fock values), the cubic crystal-field splitting is $10Dq \approx 1.4$ eV, and the hopping strengths are $V_{pd\sigma} =-2$ eV and $V_{pd\pi} = 1$ eV.

\subsection{Partial excitation density distribution}
\label{appen:PES}

To explore the intermediate and final states involved in XAS and RIXS processes, we compute the partial excitation density (PED) distribution of multiplet states with specific configurations such as $t_{2g\uparrow}^3$, $t_{2g\uparrow}^2e_{g\downarrow}$, $t_{2g\uparrow}^3e_{g\uparrow}L^1$, $t_{2g\uparrow}^3e_{g\uparrow}c^1$, and so on, where $L$ and $c$ represent ligand (bath) hole and Mn core $2p$ hole orbitals, respectively.
Let $\left| \Phi_l^\xi \right>$ denote the $l$th state for a given $\xi$ configuration. The partial excitation density $\Lambda_{\xi} \left( \omega \right)$ is calculated as follows:
\begin{equation}
\Lambda_{\xi}\left( \omega \right)  = \sum_n \sum_{l}
\left| \big< \Psi_n \big| \Phi_l^\xi \big> \right|^2
\delta \left( \omega - E_n + E_g \right) \nonumber
=-\frac{1}{\pi} \textrm{Im} \sum_l \big< \Phi_l^\xi\big|
 \frac{1}{ \omega - H + E_g + i \delta_b} \big| \Phi_l^\xi\big>,
\label{Eq:pPES}
\end{equation}
where $E_n$ and $\Psi_n$ are the energy and eigenstate of the $n$th state, respectively, and $E_g$ is the ground state energy. We set the broadening parameter $\delta_b$ to be $0.3$ eV. We utilize the Lanczos method to solve this equation.

% \section{The XMCD contribution of the CT states}
%
% To analyze the XMCD spectra, we calculate the PED distribution of the final XMCD states derived from the CT states $\big|t^3_{2g\uparrow}e_{g\uparrow}^1L_{e_g\uparrow}^1 \big>$ and $\big|t^3_{2g\uparrow}e_{g\uparrow}^2L_{e_g\uparrow}^2 \big>$, which mainly constitute the ground multiplet state, through the dipole transition operators of $D_{+}^{t_{2g}}$ and $D_{+}^{t_{2g}}$.
% Results are presented in Fig.~\ref{Fig_dstL}.

\begin{figure}[t]
\centering
\includegraphics[width= \columnwidth]{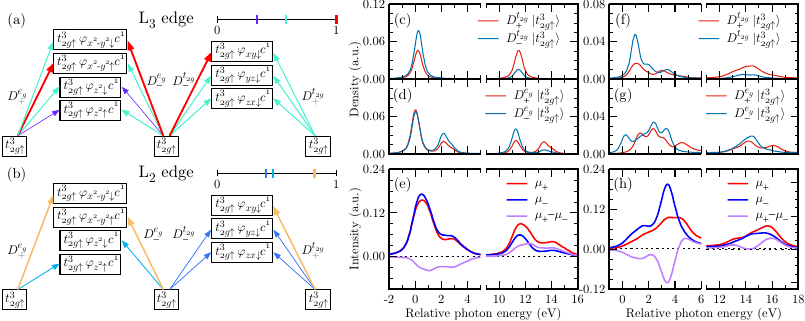}
\caption {
Schematic diagrams of dipole transitions from the $\left|t_{2g\uparrow}^3 \right>$ state at the (a) $L_3$ and (b) $L_2$ edges. $D^{t_{2g}}_{+}$ and $D^{e_g}_{+}$ ($D^{t_{2g}}_{-}$ and $D^{e_g}_{-}$) denote the dipole transition operators for right (left) circularly polarized photons from core $2p$ to valence $t_{2g}$ and $e_{g}$ orbitals, respectively.
The thickness of arrows indicates the strength of dipole transitions, with relative values marked at the top-right horizontal lines.
The partial excitation distribution (PED) behaviors of multiplet states derived from
$\big|t^3_{2g\uparrow}\big>$ by the dipole operators $D_{+}^{t_{2g}}$, $D_{-}^{t_{2g}}$, $D_{+}^{e_{g}}$, and $D_{-}^{e_{g}}$ when the scale parameter $s$ is $0$ for (c) \& (d), and $1$ for (f) \& (g).
 The theoretical $L_3$-edge X-ray magnetic circular dichroism (XMCD) spectra for (e) $s = 0$ and (h) $s=1$.
 The scale parameter is defined as $F_{pd}^2= 5.7508\times s$, $G_{pd}^1= 4.2864\times s$, and $G_{pd}^3 = 2.4382\times s$ eV.
 Other parameters are presented in Table~I of the main text.
}
\label{Fig_dst}
\end{figure}

\begin{figure}[t]
\centering
\includegraphics[width= \columnwidth]{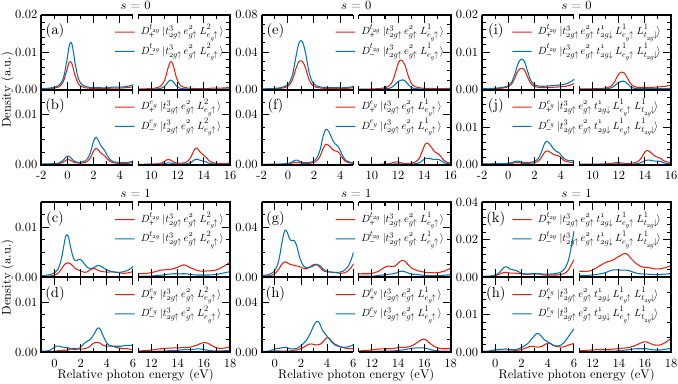}
\caption {
The partial excitation distribution behaviors of multiplet states derived from
(a)--(d) $\big|t^3_{2g\uparrow}e_{g\uparrow}^2 L_{e_g\uparrow}^2 \big>$ for Mn$^{4+}$, and (e)--(f) $\big|t^3_{2g\uparrow}e_{g\uparrow}^2L_{e_g\uparrow}^1 \big>$ and (i)--(l) $\big|t^3_{2g\uparrow}e_{g\uparrow}^2 t_{2g\downarrow} L_{e_g\uparrow}^1 L_{t_{2g}\uparrow}^1 \big>$ for Mn$^{3+}$ through the dipole transition operators of $D_{+}^{t_{2g}}$ and $D_{+}^{t_{2g}}$.
$c$, $L_{e_g\uparrow}$, and $L_{t_{2g}\downarrow}$ denote the core-hole orbital, ligand (bath) hole orbital with the $e_g$ symmetry and $\uparrow$ spin state, and ligand (bath) hole orbital with the $t_{2g}$ symmetry and $\downarrow$ spin state, respectively.
The scale parameter of core-valence exchange correlations $s$ is set by $s=0$ for (a), (b), (e), (f), (i), and (j) and $s=1$ for (c), (d), (g), (h), (k), and (l), respectively.
The scale parameter $s$ is defined as $F_{pd}^2= 5.7508\times s$, $G_{pd}^1= 4.2864\times s$, and $G_{pd}^3 = 2.4382\times s$ eV.
Colored arrows refer to the peak positions of XMCD spectra.
Other parameters are presented in Table 1 of main text.
}
\label{Fig_dstL}
\end{figure}

\begin{figure}[!h]
\centering
\includegraphics[width=0.7\columnwidth]{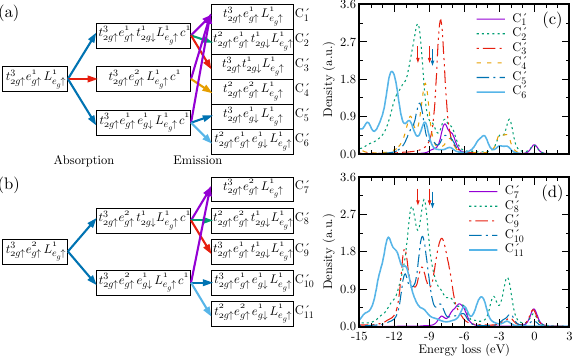}
\caption {
Schematic diagram of the resonant inelastic X-ray scattering (RIXS) process from relevant ground configurations: (a) $t_{2g\uparrow}^3 e_{g\uparrow}^1 L_{e_g\uparrow}$ and (b) $t_{2g\uparrow}^3 e_{g\uparrow}^2 L_{e_g\uparrow}$. Here, $L_{e_g\uparrow}$ denotes a ligand (bath) hole orbital with $e_g$ orbital symmetry and $\uparrow$ spin state.
The partial excitation distribution (PED) behaviors of relevant charge-transfer states in final states in the RIXS process for (a) Mn$^{4+}$ and (b) Mn$^{3+}$ calculations.
Arrows in (c) and (d) indicate the peak positions of RIXS-MCD spectra in Fig.~5(e) and (f) in the main text.
Parameters are presented in Table I of the main text.
}
\label{Fig_spes}
\end{figure}

%\bibliography{references}

\end{document}